\documentclass[twocolumn,breaklinks]{svjour3}
\pdfoutput=1
\journalname{}

\date{12 Feb 2021, last updated 16 Dec 2021}
\providecommand{\svhline}{\hline}
\makeatletter
\expandafter\let\csname opt@amsmath.sty\endcsname\relax% Remove options passed to amsmath
\AtBeginDocument{
  \mathindent=15pt % Restore \mathindent
  \@mathmargin\@centering} % Restore \@mathmargin
\makeatother

\usepackage{mathptmx}       % selects Times Roman as basic font
\usepackage{helvet}         % selects Helvetica as sans-serif font
\usepackage{courier}        % selects Courier as typewriter font
\usepackage{makeidx}         % allows index generation
\usepackage{graphicx}        % standard LaTeX graphics tool
\usepackage[bottom]{footmisc}% places footnotes at page bottom
\usepackage{textcomp}

\usepackage{amssymb,mathtools}
\renewcommand{\nRightarrow}{\,\not\!\Rightarrow}
\renewcommand{\nLeftarrow}{\,\not\!\Leftarrow}
\renewcommand{\nLeftrightarrow}{\,\not\!\Leftrightarrow}
\usepackage[shortlabels]{enumitem}
\usepackage{xcolor}
\usepackage[normalem]{ulem}
\usepackage{longtable}
\usepackage{float}

\usepackage{braket}
\newcommand{\gen}[1]{\langle #1\rangle}

\makeindex             % used for the subject index
\usepackage[style=authoryear,%phys,biblabel=brackets,
dashed=false, 
sorting=nyt,backend=biber, doi=true, url=true, isbn=true,eprint=true]{biblatex}%
\usepackage{zotero}%
\renewbibmacro{in:}{%
  \ifentrytype{article}{}{\printtext{\bibstring{in}\intitlepunct}}}
\DeclareFieldFormat[article,periodical]{volume}{\mkbibbold{#1}}
\usepackage[colorlinks=true, linkcolor=blue, urlcolor=blue,citecolor=blue]{hyperref}

\begin{document}
\sloppy
\title{Conjugate Logic}
\author{Niklas Johansson, Felix Huber, Jan-Åke Larsson}
\institute{Niklas Johansson 
\at Dept of Electrical Engineering, Linköping University, SE-581 83 Linköping, SWEDEN, \email{niklas.johansson@liu.se}
\and Felix Huber
\at Atomic Optics Department, Jagiellonian University, PL-30-348 Kraków, POLAND,
\email{felix.huber@uj.edu.pl}
\and Jan-Åke Larsson
\at Dept of Electrical Engineering, Linköping University, SE-581 83 Linköping, SWEDEN, \email{jan-ake.larsson@liu.se}
}
\maketitle

%\abstract*{}

\abstract{We propose a \textit{conjugate logic} that can capture the behavior of quantum and quantum-like systems. 
The proposal is similar to the more generic concept of epistemic logic: it encodes knowledge or perhaps more correctly, predictions about outcomes of future observations on some systems.
For a quantum system, these predictions are statements about future outcomes of measurements performed on specific degrees of freedom of the system.
The proposed logic will include propositions and their relations including connectives, but importantly also transformations between propositions on conjugate degrees of freedom of the systems. 
A key point is the addition of a transformation that allows to convert propositions about single systems into propositions about correlations between systems. 
We will see that subtle choices of the properties of the transformations lead to drastically different underlying mathematical models; one choice gives stabilizer quantum mechanics, while another choice gives Spekkens' toy theory.
This points to a crucial basic property of quantum and quantum-like systems that can be handled within the present conjugate logic by adjusting the mentioned choice.
It also enables a discussion on what behaviors are properly quantum or only quantum-like, relating to that choice and how it manifests in the system under scrutiny.
}

\section{Introduction}
\label{sec:introduction}
Due to the current fast development of technology towards quantum information processing, there is great interest in different tools for understanding the behavior of quantum systems.
To model this processing, often a digital information representation is used. 
This calls for a suitable associated logic.
Of course, there are already well-known logics to choose from \parencite[see e.g.,][and references therein]{Coecke2000}.

Quantum logic was proposed already by \textcite{Birkhoff1936}, by extending the language of standard logic to encompass Hilbert space structure. 
In their proposal, they start from the Hilbert space description itself and describe a logic that, for example, replaces logical negation with orthogonality in the Hilbert space. 
More recently, there has been a drive to avoid using any Hilbert space structure as a postulate, and to look instead at other ways of building up a mathematical structure that eventually \textit{arrives} at quantum mechanical behavior. 
This has grown into an entire field of scientific investigation \parencite[to give a few examples]{Hardy2001,Pawlowski2009a,Chiribella2011,Masanes2011}.
Part of the discussion is the difficulty to avoid a direct postulate of an underlying Hilbert space.
Here we will make yet another attempt to build up a structure that avoids inserting Hilbert space structure by hand.
We will, however, still retain propositions that relate to {phase space} as in \textcite{Birkhoff1936}, but not the direct relation used there where propositions \textit{are} subsets of phase space.

Our propositions will instead concern predictions of future measurement outcomes of conjugate degrees of freedom.
There is a connection to the concept of \textit{conjugate coding} as proposed already in the 1970's by Wiesner (but only published in \cite*{Wiesner1983}), but there is also an important difference.
Conjugate coding is about encoding data into conjugate degrees of freedom. 
Here we aim for the converse: \textit{conjugate logic} is intended to describe knowledge about the system in the form of predictions of future measurement outcomes of conjugate degrees of freedom, making this a more epistemic approach.

Epistemic logic has already been studied as a subfield of epistemology concerned with logical approaches to knowledge, belief and related notions. 
While {\em any} logic with an epistemic interpretation may be called an epistemic logic, the most widespread type of epistemic logic is that of modal logics \parencite{vonWright1951}.
Epistemic modal logic is concerned with agents, their knowledge, and beliefs. It formally encodes for example, one agent's belief about another agent's knowledge by adding knowledge predicates on top of standard logic. 
Unfortunately this is not well adapted to the quantum-mechanical situation, where so-called quantum contextuality \parencite{Kochen1967} prohibits assigning a consistent set of values to the complete set of propositions in the underlying standard logic.

For this reason, the logic we aim for here is intended to capture fundamental uncertainty within the logic itself: that statements not only have an unknown truth value, but simply do not possess any intrinsic truth value.
In this sense it is no longer a question of incomplete knowledge about an existing truth value, but there are composite propositions that are fundamentally uncertain.
The {\em conjugate logic} we propose here will allow but does not force quantum behavior, and will capture the notion of incomplete knowledge in the strongest possible sense: it will contain fundamentally uncertain composite propositions.

In this logic, propositions will concern predictions of quantum-like measurement outcomes on physical systems.
Logic propositions then discretize the measurement outcomes into true/false, in general being sentences on the form ``the measurement outcome will lie between $a$ and $b$''.
Here, we will restrict ourselves to dichotomic measurement outcomes, without loss of generality.
Such a dichotomic measurement is usually associated with a spin measurement along the $Z$ axis in three-dimensional space, forming the first coordinate in a two-dimensional discrete phase space \parencite{Wootters2003}.
The second coordinate is usually associated with a spin-$X$ measurement, and diagonal lines in the discrete phase space are associated with spin-$Y$ measurement.

Our proposed conjugate logic will contain these conjugate degrees of freedom, and importantly also transformations between them, not only on single systems but also transformations involving correlations between systems. 
This makes our approach different from other three-valued logics that attempt to capture aspects of quantum mechanics, but do not include conjugate degrees of freedom as an essential part, nor transformations between them, see e.g., \textcite{Reichenbach1944}.
In addition, the inclusion of these transformations enables a logic-language basis for stabilizer quantum mechanics complete with Clifford-group transformations \parencite{Calderbank1998,Gottesman1998,Gottesman1998a}.

Stabilizer quantum mechanics is an important tool in quantum information processing.
Our construction will be able to reproduce the behavior of this subset of quantum mechanics, but will avoid postulating an underlying Hilbert space.
It will also enable a logic-language basis for Spekkens' toy theory and extensions of it \parencite{Spekkens2007,Wallman2012,Blasiak2013,Johansson2019,Lillystone2019} that are useful in foundational considerations on these issues.
Let us now turn to the explicit construction.

\section{Conjugate logic propositions, negation, conjunction and disjunction}
\label{sec:atomic}

We start with something that looks reasonably familiar; a three-valued logic where one truth value will denote an \textit{indeterminate} outcome.
Note that we immediately deviate from standard logic by adding several conjugate {degrees of freedom} intending to capture the idea that not all measurements are compatible, meaning simultaneously predictable. 
Standard propositional logic contains {propositions}, usually denoted $p$ and $q$, that can have truth values in \{false,true\} or \{0,1\}.
In the new logic an atomic proposition is a prediction for the outcome of a measurement, that is, a statement on the form ``measurement of the $Z$ degree of freedom would give the outcome 0'', denoted $\gen Z$.
We choose ``outcome 0'' because this gives a natural correspondence to the stabilizer formalism where the notation $\gen \cdot$ is used for a \textit{stabilizer}, a transformation that keeps a quantum state unchanged.
There, $\gen Z$ is the notation for the stabilizer of the quantum state $|0\rangle$, which gives a direct connection to the proposition above.

Similarly, we add propositions to two other conjugate {degrees of freedom} whose corresponding propositions $\gen X$ and $\gen Y$ are statements on predictions about these measurement outcomes.
Propositions of this kind are atomic much like simple propositions of ordinary propositional logic, and serve as building blocks for more complex expressions below. 
Statements on measurement outcomes of different systems will be denoted by numerical indexes $Z_j$ or by position in a string of symbols.

Propositions can be true or false just as in standard propositional logic. 
They may also hold no truth value meaning that the measurement outcome is indeterminate, uncertain, unknown, or cannot be predicted, denoted ``?''.
In some presentations of propositional logic one can see ``T'' for true (1) and ``F'' for false (0), so we could add ``I'' for indeterminate \parencite{Reichenbach1944}, in what follows we will use ``?'' for readability.
Importantly, the conjugate propositional logic we construct here will contain statements that \textit{cannot} all have truth values simultaneously.
Although we will assume in what follows that at least one statement per system is allowed to have a truth value.

Just as in standard logic we can form connectives. 
In standard logic, the unary connective NOT ($\neg$) converts a proposition into its negation.
It is natural to extend this notion such that the conjugate logic NOT gives truth value 0 if the negated proposition has truth value 1, and gives 1 if the negated proposition have truth value 0; finally, it gives the indeterminate ``?'' truth value if the negated proposition is indeterminate~``?''.

Now consider the standard logic conjunction AND ($\wedge$): it gives truth value 0 if one of the constituent propositions has truth value 0, and 1 if both the constituent propositions have truth value 1.
It is natural to extend this notion so that the conjugate logic conjunction has exactly this behavior and that it gives indeterminate ``?'' otherwise, i.e., in the case when at least one constituent is indeterminate and the other is either indeterminate or 1. 
For example, we denote ``[measurement of the $Z_1$ degree of freedom would give the outcome 0] AND [measurement of the $X_2$ degree of freedom would give the outcome 0]'' by $\gen{Z_1}\wedge \gen{X_2}$; or sometimes $\gen {ZI}\wedge\gen {IX}$ or $\gen {ZI,IX}$. 

Conjugate logic disjunction (OR, $\vee$) can be defined similarly by extending the standard truth table to cases with indeterminate constituent propositions.
Thus, the conjugate logic disjunction gives truth value 1 if one of the constituent propositions has truth value 1, truth value 0 if both the constituent propositions have truth value 0, and gives indeterminate ``?'' otherwise, i.e., if one of the constituents is indeterminate and the other is either indeterminate or 0.
We see that the conjugate logic conjunction and disjunction mirror each other just as they do in standard logic (see ``de Morgan's laws'' in Table~\ref{tab:connectives}).

Finally, conjugate logic exclusive disjunction (XOR, $\veebar$) is also an extension of the standard logic truth value 1 if one of the constituent propositions has truth value 1 and the other 0, and truth value 0 if both the constituent propositions have truth value 0 or 1. We simply let it give indeterminate ``?'' if either or both of the constituents are indeterminate.

\begin{table}[b]
\caption{Conjugate logic connectives}
\label{tab:connectives}       % Give a unique label
\centering
\begin{tabular}{@{\hspace{3mm}}c@{\hspace{4mm}}c@{\hspace{9mm}}c@{\hspace{9mm}}c
@{\hspace{3mm}}c@{\hspace{3mm}}c@{\hspace{9mm}}c@{\hspace{2mm}}c}
\hline
\noalign{\smallskip}
$p$&$q$&$\neg p$&$p\wedge q$&$p\vee q$
&$p \veebar q$&$p\rightarrow q$&$p\leftrightarrow q$\\
\noalign{\smallskip}
\svhline
\noalign{\smallskip}
0&0&1&0&0&0&1&1\\
0&?&1&0&?&?&1&0\\
0&1&1&0&1&1&1&0\\
?&0&?&0&?&?&0&0\\
?&?&?&?&?&?&1&1\\
?&1&?&?&1&?&1&0\\
1&0&0&0&1&1&0&0\\
1&?&0&?&1&?&0&0\\
1&1&0&1&1&0&1&1\\
\noalign{\smallskip}
\hline
\noalign{\smallskip}
\end{tabular}
\end{table}
\begin{table}[b]
\caption{Truth table for the inverse law}
\label{tab:inverselaw}       % Give a unique label
\centering
\begin{tabular}[t]{@{\hspace{4mm}}c@{\hspace{4mm}}c@{\hspace{10mm}}c@{\hspace{3mm}}c
@{\hspace{10mm}}c@{\hspace{2mm}}
c@{\hspace{2mm}}}
\hline
\noalign{\smallskip}
$p$&$\neg p$&$p\vee\neg p$&$\gen I$&$p\wedge\neg p$&$\gen{\neg I}$\\
\noalign{\smallskip}
\svhline
\noalign{\smallskip}
0&1&1&1&0&0\\
1&0&1&1&0&0\\
?&?&?&1&?&0\\
\noalign{\smallskip}
\hline
\noalign{\smallskip}
\end{tabular}
\end{table}

\section{Conjugate logic material and logical conditionals, tautology and contradiction}
\label{sec:conditionals}

We have now come to the conjugate logic \textit{material conditional} and \textit{material biconditional}~($\rightarrow$ and $\leftrightarrow$), where the qualifier ``material'' is used to distinguish the connective ($\rightarrow$) from the ``formal'' conditional~($\Rightarrow$) as \textcite{Russell1903} writes,
or the ``logical'' conditional as it is now more commonly known.
Perhaps jokingly, one could add that a better name would be \textit{immaterial conditional} instead of \textit{material conditional} when describing quantum and quantum-like systems. We will not press the issue more here, after all, \textit{quantum systems are the dreams that stuff is made of}.

To construct material conditional and biconditional that incorporates an intrinsic uncertainty, we will take a closer look at two alternatives.
One is to construct conditionals such that they follow identities from ordinary propositional logic. 
For example, then $p\leftrightarrow q$ would be identical to $(p\wedge q)\vee(\neg p\wedge\neg q)$. 
However, this does not capture the desired behavior well: the latter expression gives an indeterminate value ``?'' if both $p$ and $q$ are indeterminate, while the material biconditional should compare if the truth values are equal.
Thus, in our view, a better alternative is to have the material biconditional (EQUIVALENT TO, $\leftrightarrow$) compare the two constituent propositions and have $p\leftrightarrow q$ be 1 if $p$ and $q$ have the same truth value, and 0 otherwise.
In particular, $p\leftrightarrow q$ is 1 if both $p$ and $q$ are indeterminate. 
The logical biconditional between two composite propositions $p\Leftrightarrow q$ can now be defined as usual, namely as the situation in which $p\leftrightarrow q$ is always true.
This now coincides with the statement that $p$ and $q$ have identical truth table entries.

The material conditional (IMPLIES, $\rightarrow$) can similarly be extended to have $p\rightarrow q$ be 1 if $p$ is 0, if $q$ is 1, or (in addition to the standard definition) if $p$ and $q$ have the same truth value, and 0 otherwise.
The additional clause captures the notion of implication in the case when both $p$ and $q$ are indeterminate, encoding that when $p$ is indeterminate we cannot draw any conclusion about $q$ so that an indeterminate truth value is acceptable by the connective. 
This completes the list of connectives in Table~\ref{tab:connectives}, and allows us to define the logical conditional between two composite propositions $p\Rightarrow q$ as the situation in which $p\rightarrow q$ is always true.

Using the notation $\gen{\neg Z}$ for the statement ``measurement of the $Z$ degree of freedom would give the outcome 1'' we note that
\begin{equation}
  \neg\gen Z\Leftrightarrow\gen{\neg Z},
\end{equation} 
where the equivalence holds also for indeterminate truth values.
The propositions $\gen I$ and $\gen {\neg I}$ represent tautology and contradiction, respectively, corresponding to trivial measurements that always have the outcome 0 or 1.

\section{Equivalence relations and implication laws}

In standard logic there are a number of logical equivalence relations and implication laws. 
Some of these are important properties of logic expressions like reflexivity, symmetry, and transitivity, and these are retained by conjugate logic as defined here.
However, some equivalences that are common tools in standard propositional logic no longer hold.
The most prominent example is the ``inverse law,'' which in standard propositional logic tells you that ($p\vee\neg p$) is a tautology and that ($p\wedge\neg p$) is a contradiction. 
These relations fail because of the indeterminate values involved, see the relevant truth tables in Table~\ref{tab:inverselaw}.
It follows that seemingly innocuous simplifications are no longer available to us, for example,
\begin{equation}
\gen{ZI,IX}\vee\gen{\neg ZI,IX}
\Leftrightarrow
\Big(\gen{ZI}\vee\gen{\neg ZI}\Big)\wedge \gen{IX}
\,\not\!\Leftrightarrow
\gen{IX}.\label{eq:noinverselaw}
\end{equation}

It is a simple exercise to verify that for generic propositions $p$, $q$, and $r$, the following equivalences hold, and do not hold (see Appendix~\ref{sec:tables} for details).\medskip

\noindent
\begin{tabular}[l]{@{\hspace{0mm}}r@{\hspace{1mm}}l@{\hspace{1mm}}l@{\hspace{1mm}}}
(E1)&\textbf{Double negation}&$\neg\neg p\Leftrightarrow p$\\ 
(E2)&\textbf{De Morgan's laws}&$\neg(p\wedge q)\Leftrightarrow \neg p\vee \neg q$\\ 
&&$\neg(p\vee q)\Leftrightarrow \neg p\wedge \neg q$\\ 
(E3)&\textbf{Commutative laws}&$p\wedge q\Leftrightarrow q\wedge p$\\ 
&&$p\vee q\Leftrightarrow q\vee p$\\ 
(E4)&\textbf{Associative laws}&$p\wedge(q\wedge r)\Leftrightarrow (p\wedge q)\wedge r$\\ 
&&$p\vee (q\vee r)\Leftrightarrow (p\vee q)\vee r$\\ 
(E5)&\textbf{Distributive laws}&$p\wedge(q\vee r)\Leftrightarrow (p\wedge q)\vee(p\wedge r)$\\ 
&&$p\vee (q\wedge r)\Leftrightarrow (p\vee q)\wedge(p\vee r)$\\ 
(E6)&\textbf{Idempotence}&$p\wedge p\Leftrightarrow p$\\ 
&&$p\vee p\Leftrightarrow p$\\ 
(E7)&\textbf{Identity laws}&$p\wedge \gen{I}\Leftrightarrow p$\\ 
&&$p\vee \gen{\neg I}\Leftrightarrow p$\\ 
(E8)&\textbf{Domination laws}&$p\wedge \gen{\neg I}\Leftrightarrow \gen{\neg I}$\\ 
&&$p\vee \gen{I}\Leftrightarrow \gen{I}$\\ 
(E9)&\sout{Inverse laws}&$p\wedge \neg p\nLeftrightarrow \gen{\neg I}$\\ 
&&$p\vee \neg p \nLeftrightarrow \gen{I}$\\ 
(E10)&\textbf{Absorption laws}&$p\wedge (p\vee q)\Leftrightarrow p$\\ 
&&$p\vee (p\wedge q)\Leftrightarrow p$\\ 
(E11)&\sout{Implication law}&$p\rightarrow q \nLeftrightarrow \neg p\vee q$\\ 
(E12)&\textbf{Contrapositive 
 law}&$p\rightarrow q \Leftrightarrow \neg p\rightarrow \neg q$\\ 
(E13)&\textbf{Equivalence 
 law}&$p\leftrightarrow q \Leftrightarrow (p\rightarrow q)\wedge(q\rightarrow p)$\\ 
\end{tabular}\medskip

An example of how the inverse law breaks in quantum mechanics is the double slit experiment, where the proposition $p=$ ``the particle passes through the right-hand slit'' corresponds to a dichotomic measurement.
If interference is desired from the experiment, it must be arranged so that $p$ is indeterminate ``?'', in which case $p\wedge\neg p$ is indeterminate ``?'' so not a contradiction, and $p\vee\neg p$ is indeterminate  ``?'' so not a tautology.
We find that this describes the situation better than the popular-science ``the particle passes through both of the slits (both $p$ and $\neg p$ are true so $p\wedge\neg p$ is true) and simultaneously none of the slits (both $p$ and $\neg p$ are false so $p\vee\neg p$ is false),'' which at best is just confusing.

For the inverse laws only the trivial implications remain, while the implication law does not hold in either direction. 
A similar exercise for logical implications gives the following list, details in Appendix~\ref{sec:tables}.\medskip

\noindent
\begin{tabular}[l]{@{\hspace{0mm}}r@{\hspace{1mm}}l@{\hspace{1mm}}l@{\hspace{1mm}}}
(I1)&\textbf{Modus ponens}&$(p\rightarrow q)\wedge p\Rightarrow q$\\
(I2)&\textbf{Law of syllogism}&$(p\rightarrow q)\wedge (q\rightarrow r)\Rightarrow (p\rightarrow r)$\\
(I3)&\textbf{Modus tollens}&$(p\rightarrow q)\wedge \neg q\Rightarrow \neg p$\\
(I4)&\textbf{Conjunctive simpl.}&$p\wedge q\Rightarrow p$\\
(I5)&\textbf{Disjunctive ampl.}&$p\Rightarrow p\vee q$\\
(I6)&\sout{Disjunctive syllogism}&$(p\vee q)\wedge \neg q\nRightarrow p$\\
(I7)&\textbf{Proof by contradiction}&$(\neg p\rightarrow \gen{\neg I})\Rightarrow p$\\
(I8)&\textbf{Proof by cases}&\small$(p\rightarrow r)\wedge (q\rightarrow r)\Rightarrow (p\vee q)\rightarrow r$\\
\end{tabular}\medskip

Most of the standard rules that we use when proving theorems do still hold, the exception is disjunctive syllogism, that succumbs to the same problem as the implication law: when $q$ is indeterminate, we cannot draw a conclusion about $p$. Note that modus tollens still holds, encompassing a slightly stronger requirement.

\section{Fundamental uncertainty}

We now arrive at a crucial point in the construction, the very reason to include the indeterminate value for the propositions used. 
So far, the indeterminate value could correspond to lack of knowledge about the ``actual'' ontic (existing) value of the property being measured. 
But when making statements about quantum systems, one should take into account that the standard mathematical description does not contain such ontic values, but rather, only allows calculation of probabilities after specifying which property is to be measured.
The debate goes back to the founding fathers of quantum theory \parencite{Einstein1935,Bohr1935}.
We do not wish to take a stance on that particular issue here, merely describe a truly epistemic logic, that encompasses the possibility that indeterminate values ``?'' not just denote lack of knowledge but may be \textit{fundamentally uncertain}.

In quantum mechanics, not all measurement outcomes can simultaneously be predicted with certainty (probability~1 as EPR~\cite*{Einstein1935} write). 
Also in quantum-like systems, like Spekkens' toy theory, not all measurement outcomes can be simultaneously predicted with certainty.
It is then natural to require that in the present conjugate logic that not all propositions can hold truth values simultaneously. 
In particular, the uncertainty principle holds in the form of a bound on the descriptional power of this conjugate propositional logic concerning atomic propositions, i.e., propositions on single degrees of freedom of single systems.\medskip

\begin{center}
\fbox{\parbox{.8\linewidth}{\textbf{Postulate: Bound on descriptional power}

No more than a single atomic proposition can be true or false for any single system.
}}
\end{center}\medskip

We can immediately conclude that some composite propositions on a single system cannot have a definite truth value, 
for example,
\begin{equation}
  \gen Z\veebar\gen{X}\Leftrightarrow\ ?\ .
\end{equation} 
This bound on descriptional power corresponds directly to the uncertainty relation in quantum mechanics \parencite{Heisenberg1927} and to the knowledge balance principle of Spekkens' toy theory \parencite{Spekkens2007}.
A set of propositions that {\em can} have truth values simultaneously we will call \textit{compatible}. 
Since not all propositions can have simultaneous truth values, not all measurements are compatible. 

The postulate has consequences on what we can say about systems in general, for example about correlations between systems. 
However, to arrive at a precise statement we will need transformations between propositions; these correspond to physical operations performed on the physical systems that we study.

\section{Transformations of propositions}

Our propositions concern predicted outcomes of measurements performed on physical systems.
As such, one of our systems can be subjected to a range of physical transformations, the simplest case is 180\textdegree\ rotation of the system around one of the conjugate coordinate axes.
Such a rotation will conserve the proposition that concerns the axis, but perform a negation of the other two.
It is therefore natural to use the conserved axis as label, in complete parallel with the stabilizer formalism:
\begin{alignat}6
  \varphi_X \gen X &\Leftrightarrow\;& \gen {X},\phantom{\neg}\quad
  \varphi_X \gen Y &\Leftrightarrow\;& \gen {\neg Y},\quad
  \varphi_X \gen Z &\Leftrightarrow\;& \gen {\neg Z},\\
  \varphi_Y \gen X &\Leftrightarrow\;& \gen {\neg X},\quad
  \varphi_Y \gen Y &\Leftrightarrow\;& \gen {Y},\phantom{\neg}\quad
  \varphi_Y \gen Z &\Leftrightarrow\;& \gen {\neg Z},\\
  \varphi_Z \gen X &\Leftrightarrow\;& \gen {\neg X},\quad
  \varphi_Z \gen Y &\Leftrightarrow\;& \gen {\neg Y},\quad
  \varphi_Z \gen Z &\Leftrightarrow\;& \gen {Z}.\phantom{\neg}
\end{alignat}
Since these are equivalence relations, the transformations preserve truth values and compatibility relations.
The ``Phase rotation''~$S$ comes in two variants.
The quantum phase rotation corresponds to a 90\textdegree\ rotation around the $Z$ axis, so that
\begin{subequations}
\begin{alignat}6
    \varphi_S \gen X &\Leftrightarrow\;& \gen Y,\quad 
    \varphi_S \gen Y &\Leftrightarrow\;& \gen {\neg X},\quad 
    \varphi_S \gen Z &\Leftrightarrow\;& \gen Z.
    \label{eq:S}
\end{alignat}
An alternative transformation is used in Spekkens' toy theory~\parencite{Pusey2012} where the rotation is followed by an inversion along the $Z$ axis,
\begin{alignat}6
    \varphi_S \gen X &\Leftrightarrow\;& \gen Y,\quad 
    \varphi_S \gen Y &\Leftrightarrow\;& \gen {\neg X},\quad 
    \varphi_S \gen Z &\Leftrightarrow\;& \gen {\neg Z}.
    \label{eq:toyS}
\end{alignat}
\end{subequations}
The effect of this difference is small, note that in both cases $\varphi_S\varphi_S=\varphi_Z$.

We now arrive at a crucial step in the construction, the ``Hadamard'' transformation.
This transformation also comes in two variants with a seemingly small difference in the transformation itself, but this difference will instead have very important consequences for the type of model that can be used to describe the system, we will expand on this below.
The quantum-mechanical Hadamard corresponds to a physical rotation around an axis 45\textdegree\ between $X$ and $Z$, giving here a transformation that interchanges $X$ and $Z$ and inverts $Y$, 
\begin{subequations}
\begin{alignat}6
    \varphi_H \gen X &\Leftrightarrow\;& \gen Z,\quad
    \varphi_H \gen Y &\Leftrightarrow\;& \gen {\neg Y},\quad
    \varphi_H \gen Z &\Leftrightarrow\;& \gen X.
    \label{eq:Hadamard}
\end{alignat}
The alternative Hadamard transformation~\parencite{Pusey2012} used in Spekkens' toy theory corresponds to an mirror operation over the plane spanned by the mentioned axis 45\textdegree\ between $X$ and $Z$ and the $Y$ axis, i.e., a transformation that interchanges $X$ and $Z$ and preserves $Y$, 
\begin{alignat}6
    \varphi_H \gen X &\Leftrightarrow\;& \gen Z,\quad
    \varphi_H \gen Y &\Leftrightarrow\;& \gen {Y},\quad
    \varphi_H \gen Z &\Leftrightarrow\;& \gen X.
    \label{eq:toyHadamard}
\end{alignat}
\end{subequations}
Finally, the identity transformation $\varphi_I$ leaves all propositions unchanged.
For both versions of the Phase rotation and Hadamard, we arrive at a noncommutative group generated by $\varphi_S$ and $\varphi_H$, since
\begin{equation}
\varphi_S\varphi_S=\varphi_Z,\quad\varphi_H\varphi_Z\varphi_H=\varphi_X\quad\text{and}\quad \varphi_S\varphi_X\varphi_S^{-1}=\varphi_Y. 
\end{equation} 
This also generates all single-system propositions from one, say $\gen Z$.

\section{Composite system propositions and transformations}

For composite systems, there remains to handle joint measurements, which enable statements about correlation between measurement outcomes without making statements about the individual measurement outcomes.
The statement ``a joint measurement of the XOR between $Z_1$ and $Z_2$ would give outcome 0'' can be written $\gen{Z_1\veebar Z_2}$, or to conform with standard stabilizer notation $\gen{ZZ}$, we will use the latter notation below.
 
To avoid confusion please note that $\gen{ZZ}$ corresponds to a joint measurement of a single dichotomic value that gives the XOR, it does not correspond to individual measurement of two dichotomic values followed by calculation of the XOR between them, this would be denoted $\gen{ZI}\veebar\gen{IZ}$. 
These are different procedures, and have different consequences, in particular
\begin{equation}
\gen{ZI}\veebar\gen{IZ} \Rightarrow\gen{ZZ}\quad\text{but}\quad
\gen{ZI}\veebar\gen{IZ} \nLeftarrow\gen{ZZ}.
\end{equation}
There are some simple equivalences since the conjunction of a statement about a single system and a statement about its correlation to another system, is equivalent to a conjunction of individual statements about the two systems, for example,
\begin{equation}
\gen{XI,XX} \Leftrightarrow\gen{XI,IX}, \quad
\gen{\neg XI,XX}\Leftrightarrow\gen{\neg XI, \neg IX}.\label{eq:Xreduction}
\end{equation}

This simplification applies to conjunctions. We will now add another transformation that maps a \textit{single} proposition on the outcome of a correlation measurement to a proposition on some outcome of a single-system measurement. 
One natural choice is the $CZ$ (``controlled-$Z$''), because its action is symmetric on both subsystems, and can be summarized as
\begin{equation}
\begin{split}
  \varphi_{CZ}\gen{IX}\Leftrightarrow\gen{ZX},\;
  \varphi_{CZ}\gen{IY}&\Leftrightarrow\gen{ZY},\;
  \varphi_{CZ}\gen{IZ}\Leftrightarrow\gen{IZ},\\
  \varphi_{CZ}\gen{XI}\Leftrightarrow\gen{XZ},\;
  \varphi_{CZ}\gen{YI}&\Leftrightarrow\gen{YZ},\;
  \varphi_{CZ}\gen{ZI}\Leftrightarrow\gen{ZI},\\
  \varphi_{CZ}\gen{ZZ}\Leftrightarrow\gen{ZZ},\;
  \varphi_{CZ}\gen{XX}&\Leftrightarrow\gen{YY},\;
  \varphi_{CZ}\gen{XY}\Leftrightarrow\gen{\neg YX}.
  \label{eq:CZ}
\end{split}
\end{equation} 
The motivation for this definition (which \eqref{eq:CZ} should be viewed as) is the  standard gate-based description of a $CZ$, that applies a $\varphi_Z$ transformation on the second system if ``measurement of the $Z_1$ degree of freedom would give the outcome 1'', and $\varphi_I$ if ``measurement of the $Z_1$ degree of freedom would give the outcome 0''. 
It is immediate that $\gen{II}$, $\gen{IZ}$, $\gen{ZI}$, and $\gen{ZZ}$ are unaffected by $\varphi_Z$ on either system.
To derive the effect of this map on the proposition ``measurement of $X_2$ would give the outcome 0,'' $\gen{IX}$, we have
\begin{equation}
\begin{cases}
\;\;\varphi_{CZ}\gen{ZI,IX}
\Leftrightarrow\;\,\varphi_{II}\gen{ZI,IX}\;\,
\Leftrightarrow\gen{ZI,IX},\\
\varphi_{CZ}\gen{\neg ZI,IX}
\Leftrightarrow\varphi_{IZ}\gen{\neg ZI,IX}
\Leftrightarrow\gen{\neg ZI,\neg IX}.
\end{cases}
\end{equation}

Since we are looking for the unique statement that $\gen{IX}$ maps to, we use the equivalences in Eqn.~\eqref{eq:Xreduction} to obtain
\begin{equation}
\begin{cases}
\;\;\varphi_{CZ}\gen{ZI,IX}\Leftrightarrow\gen{ZI,ZX},\\
\varphi_{CZ}\gen{\neg ZI,IX}\Leftrightarrow\gen{\neg ZI,ZX},
\end{cases}
\end{equation}
so that, for each measurement outcome of $Z_1$ the map transforms ``measurement of $X_2$ would give the outcome 0'' into ``measurement of the XOR between $Z_1$ and $X_2$ would give the outcome~0''.
Even though the ``inverse law'' is not available, which would have directly given us $\varphi_{CZ}\gen{IX}\Leftrightarrow\gen{ZX}$, this is the only remaining possibility for a well-defined map $\varphi_{CZ}$.
Similar reasoning and the symmetry of $CZ$ gives the transformation output of the remaining single-system propositions of \eqref{eq:CZ}. 

Some further elaboration is needed for the two final entries in the table.
These can be derived from the transformation property that compatible propositions are transformed into compatible propositions, using only three of the just established single-system transformation outputs.
We know that $\gen{IX}$ and $\gen{XI}$ are compatible, and therefore $\gen{ZX}\Leftrightarrow\varphi_{CZ}\gen{IX}$ and $\gen{XZ}\Leftrightarrow\varphi_{CZ}\gen{XI}$ are compatible. 
Conversely, $\gen{IX}$ and $\gen{IY}$ are incompatible, and therefore $\gen{ZX}\Leftrightarrow\varphi_{CZ}\gen{IX}$ and $\gen{ZY}\Leftrightarrow\varphi_{CZ}\gen{IY}$ are incompatible. 
It is now possible to use single-system transformations to deduce whether a given pair of two-system propositions are compatible or not.
For example, $\gen{ZX}$ is compatible with $\gen{ZI}$, $\gen{IX}$, $\gen{XZ}$,  $\gen{YY}$, $\gen{XY}$, $\gen{YZ}$, and no other two-system propositions.
It is of particular interest that the pair \{$\gen{ZX}$,$\gen{XZ}$\} is only compatible with $\gen{YY}$ (and $\gen{II}$), which implies that the set $\{\gen{IX},\gen{XI},\gen{XX}\}$ of three pairwise compatible propositions must be transformed into the set $\{\gen{ZX},\gen{XZ},\gen{YY}\}$ or possibly $\{\gen{ZX},\gen{XZ},\gen{\neg YY}\}$.
The latter choice will give inconsistencies, for details see Appendix~\ref{sec:inconsistency}.

Both quantum mechanics and Spekkens' toy theory use the choice 
\begin{equation}
\varphi_{CZ} \gen{XX} \Leftrightarrow \gen{YY}.
\end{equation}
Then, the identity $\varphi_S\varphi_Z=\varphi_S\varphi_S\varphi_S=\varphi_Z\varphi_S$ fixes 
\begin{equation}
\varphi_{CZ} \gen{XY} 
\Leftrightarrow \varphi_{CZ}\varphi_{IS} \gen{XX} 
%\Leftrightarrow \varphi_{IS}\varphi_{CZ} \gen{XX} 
\Leftrightarrow \varphi_{IS}\gen{YY}
\Leftrightarrow \gen{\neg YX}
\end{equation}
This finishes the construction of $\varphi_{CZ}$, and enables generating the whole Clifford group, e.g., $\varphi_{CNOT}=\varphi_{IH}\varphi_{CZ}\varphi_{IH}$.
Furthermore, we can now reproduce the behavior of stabilizer quantum mechanics if the choices of phase rotation and Hadamard are made as in Eqn.~\eqref{eq:S} and~\eqref{eq:Hadamard};
we reproduce the behavior of Spekkens' toy theory if the choices are as in Eqn.~\eqref{eq:toyS} and~\eqref{eq:toyHadamard}.

\section{Clifford reduction}
\label{sec:cliffordreduction}

To transform any non-``identity'' proposition on a many-system correlation into a single-system proposition, it is enough to follow these steps.
\begin{enumerate}[(i)]
 \item 
WLOG there is a nontrivial letter at the first position.
Apply $\varphi_{S}$ (transforms $Y$ to $\neg X$) and
$\varphi_{H}$ (transforms $X$ to $Z$)
to transform the first position to $X$ and the following positions to identity $I$ or $Z$.
\item Then use $\varphi_{CZ}$ repeatedly to reduce the last $n-1$ positions containing $Z$ to $I$, to create a string with a single $X$ at the first position. 
\end{enumerate} 
We will call this a \textit{Clifford reduction} of a joint measurement. As an example,
\begin{align}
 \varphi_{ISIIIS} \gen{XYZIZY} 
 &\Leftrightarrow \gen{XXZIZX} 
 \nonumber\\
 \varphi_{IHIIIH} \gen{XXZIZX} 
 &\Leftrightarrow \gen{XZZIZZ} 
 \nonumber\\
 \varphi_{CZ_{12}}
 \varphi_{CZ_{13}}
 \varphi_{CZ_{15}}
 \varphi_{CZ_{16}} 
 \gen{XZZIZZ} &\Leftrightarrow \gen{XIIIII}\,.
\end{align}

A simultaneous Clifford reduction of two propositions into single-system propositions can also be performed. 
\begin{enumerate}[(i)]
 \item  Reduce the first proposition and perform the same transformations on the second. 
        The first proposition now reads $\gen{XI\ldots I}$.
        Either the second proposition has also been reduced to the form $\gen{\cdot I\ldots I}$ in which case we are done, or WLOG there is a nontrivial letter at the second position. 
 \item  Reduce the second proposition excluding the very first index, to put it on the form $\gen{\cdot XI\ldots I}$.
 \item  If the first index reads $I$, we have two compatible single-system expressions $\gen{XII\ldots I}$ and $\gen{IXI\ldots I}$.
        If the first index reads $X$, the two propositions can be reduced to compatible single-system propositions using Eqn.~\eqref{eq:Xreduction}.
        If the first index reads $Y$ or $Z$, the two propositions can be reduced to two incompatible single-system propositions, using Hadamards on both systems followed by $CZ$.
        Then we are done, and the two propositions have been reduced to either compatible or incompatible single-system propositions.
\end{enumerate}

We have now shown that two propositions are always simultaneously reducible to one-system propositions. 
The reduction is to two compatible propositions (for different systems) if and only if the two original propositions are compatible (but not equivalent), and otherwise the reduction gives two incompatible propositions (for the same system).

The previous discussion shows that if two propositions are compatible, then they can be simultaneously reduced to the form $\gen{XII\dots I}$ and $\gen{IXI\dots I}$.
The same is true for $n$ compatible propositions: they can be simultaneously reduced to contain one $X$ and identities $I$ on the remaining positions.
This is proven by induction as follows.
Suppose that $m$ compatible propositions have been reduced to one-index $X$'s on the first $m$ systems using a simultaneous Clifford reduction.
If $m<n$, consider an additional proposition $P_{m+1}$ compatible with the previously reduced propositions, to which the the same Clifford operations have been applied as used to reduce the $m$ first propositions.
Since $P_{m+1}$ is compatible with the previously reduced propositions, all indices $i\le m$ only contain $I$ or $X$, because otherwise $P_{m+1}$ would be incompatible with some $P_i$, $i\le m$.
The indices $i\le m$ that contain $X$ can now be converted into $I$ using Eqn.~\eqref{eq:Xreduction}.
Now reduce $P_{m+1}$ excluding the $m$ first indices, putting it on the form $\gen{I\ldots IXI\ldots I}$. 
This gives us  a set of $m+1$ compatible single-system propositions.
If $m+1=n$ we are done, otherwise repeat the process.

If the string length equals $n$ the last step is trivial, reducing $P_n$ excluding the first $n-1$ indices.
It follows that the maximal number of compatible indepdendent propositions is equal to the string length, since any single system allows for the truth value of at most one proposition.
Furthermore, any set of compatible propositions can be augmented to size equal to string length by performing simultaneous Clifford reduction, and then specifying thruth values for single systems whose truth values are yet undefined.

\section{Predictions}
\label{sec:predictions}

We have already seen that some conjunctions are equivalent to other conjunctions.
This also enables us to make predictions.
The equivalence in Eqn.~\eqref{eq:Xreduction} is the basic tool, for example, given the proposition $\gen{XI,IX}$, the measurement of $\gen{XX}$ will lead to no new knowledge; it can be obtained from the conjunction $\gen{XI, IX}$ already. It is also true that
\begin{equation}
\gen{XZ,ZX}\Leftrightarrow\varphi_{CZ}\gen{XI,IX}\Rightarrow\varphi_{CZ}\gen{XX}\Leftrightarrow\gen{YY}
\label{eq:YYprediction}
\end{equation}
which is perhaps less straightforward to see by inspection.
There is a direct link to the stabilizer formalism but we will not comment more on that link here. 

We will call a logical consequence of a conjunction of propositions a {\em prediction}.
In general, to obtain all predictions from a collection of propositions, one would need to perform a joint Clifford reduction, generate the set of predictions by repeated use of Eqn.~\eqref{eq:Xreduction}, and then invert the Clifford reduction to perform a Clifford \textit{expansion}. 
This will restore the initial collection and create the full set of predictions.
As a consequence, a collection of $n$ systems allows for at most $2^n$ propositions on single systems or correlations to be simultaneously true.
To see why, recall that at most $n$ independent single-system propositions can be simultaneously true. 
Every pair of single-system or correlation propositions yields a new single-system or correlation prediction; thus one arrives at in total $2^n$ propositions.

% 
% 1. If we measure something that we already predict, nothing changes.
% 
% 2. If we measure something compatible with what we already predict,
% we may gain knowledge.
% 
% (Given the proposition <ZI>, measuring the compatible ZX, we gain
% <ZX> or possibly <not ZX>. We now know <ZI,ZX> or <ZI,not ZX>)
% 
% 3. If we measure something incompatible with what we already predict,
% we both lose and gain knowledge.
% 
% (Given the proposition <ZI>, measuring the incompatible XZ, we gain
% <XZ> or <not XZ> but cannot anymore assign a definite value to <ZI>.
% We now know only <XZ> or <not XZ>.)

\section{Measurements affect propositions}

For a single system, no more than a single atomic proposition can be true or false at the same time.
For multiple systems this generalizes to the statement that only mutually compatible propositions can simultaneously be known:
at most $2^n$ general propositions of which there are $n$ independent ones.

We thus need to specify what happens when some degree of freedom is measured that belongs to an incompatible proposition. 
Here we take the approach that any measurement must yield a definite outcome, and that measurement enables prediction of subsequent measurement outcomes. 
Given that there is a descriptional power bound, the question is how to incorporate this new prediction.
It is reasonable that any previous proposition that is compatible with the new one remains, whereas incompatible propositions cannot coexist with the new proposition and must therefore become indeterminate.

More formally, suppose that the conjunction $P=p_1\wedge \dots \wedge p_m$ holds.
Now measure some degree of freedom of the system.
Since by our assumption the outcome enables prediction of a future measurement outcome, we capture that in a proposition $q$. 
Denote by 
\begin{equation}
 P_q = \bigwedge_{p_i\text{ compatible with } q}p_i\,.
\end{equation}
After measurement, we have the conjunction
\begin{equation}
 P_q\wedge q\,,
\end{equation}
and all predictions possible from it.
Note that if the measurement gives a proposition $q$ that is a prediction of $P$, nothing changes.
This captures the behavior of both quantum mechanics and Spekkens' toy theory.
We arrive at the following consequence of the postulate of bounded descriptional power:

\begin{center}
\fbox{\parbox{0.8\linewidth}{\textbf{Consequence: Action of Measurements}

A measurement determines the truth value of a proposition of the system and renders all incompatible propositions indeterminate.
}}
\end{center}

\section{Contextuality, and noncontextuality}
\label{sec:contextuality}

The conjugate logic constructed here allows treatment of quantum and quantum-like systems within the same formalism, capturing not only fundamental uncertainty, but also transformations between different degrees of freedom.
These transformations are needed for the emergence of properly quantum behavior, in particular transformations from knowledge on one system into knowledge about correlation between systems.
It is the addition of the latter type of transformation that makes stabilizer quantum mechanics accessible, but it is in fact one of the single-system transformations that differentiate between properly quantum, and merely quantum-like, behavior.
It also makes our three-valued logic more restrictive. 
To see this we need to consider a properly quantum phenomenon, here we will use that of quantum contextuality of the Peres-Mermin square \parencite[PM, ][]{Peres1993,Mermin1993}:
\begin{equation}
\begin{array}{ccc}
\gen{ZI}&\gen{IZ}&\gen{ZZ}\\
\gen{IX}&\gen{XI}&\gen{XX}\\
\gen{ZX}&\gen{XZ}&\gen{YY}\makebox[0pt]{\;.}
\end{array}
\label{eq:PM}
\end{equation}
The rows and columns in this square consist of compatible propositions.
In fact, we can make predictions for the last item in rows and columns from the preceding items, using Eqns.~\eqref{eq:Xreduction} and~\eqref{eq:YYprediction},
\begin{eqnarray}
&\gen{ZI}\wedge\gen{IZ}\Rightarrow\gen{ZZ};\;
\gen{IX}\wedge\gen{XI}\Rightarrow\gen{XX};&\notag\\
&\gen{ZI}\wedge\gen{IX}\Rightarrow\gen{ZX};\;
\gen{IZ}\wedge\gen{XI}\Rightarrow\gen{XZ};&\notag\\
&\gen{ZX}\wedge\gen{XZ}\Rightarrow\gen{YY}.&
\end{eqnarray}
\begin{subequations}
The exception is the final column where the derivation of the prediction (see Sec.~\ref{sec:predictions}) involves a Hadamard. 
The two different Hadamards give different predictions.
The quantum-mechanical Hadamard gives
\begin{equation}
\gen{XX,ZZ}\Leftrightarrow\varphi_{HI}\gen{ZX,XZ}\Rightarrow\varphi_{HI}\gen{YY}\Leftrightarrow\gen{\neg YY},
\label{eq:YYpredictiona}
\end{equation}
while Spekkens' toy theory Hadamard instead gives
\begin{equation}
\gen{XX,ZZ}\Leftrightarrow\varphi_{HI}\gen{ZX,XZ}\Rightarrow\varphi_{HI}\gen{YY}\Leftrightarrow\gen{YY}.
\label{eq:YYpredictionb}
\end{equation}
\end{subequations}
The latter prediction \eqref{eq:YYpredictionb} allows simultaneous assignment of truth values (other than indeterminate ``?'') to all the propositions in the PM square \eqref{eq:PM}.
The former prediction \eqref{eq:YYpredictiona} does not.

This is known as \textit{quantum contextuality} \parencite{Kochen1967,}, where the word ``context'' here refers to which conjunction a proposition is contained in, e.g., row or column in the PM square.
For a system with the quantum-mechanical Hadamard transformation, one needs to accept one of the following alternatives
\begin{enumerate}[(i)]
\item some proposition in the PM square \eqref{eq:PM} must be indeterminate ``?'',
\item some proposition in the PM square \eqref{eq:PM} must possess different values in different contexts.
\end{enumerate}
The former alternative captures fundamental uncertainty, while the latter alternative would give a \textit{contextual} ontic model.
Spekkens' toy theory, on the other hand, is a \textit{noncontextual} ontic model where propositions (i$'$) are determinate and (ii$'$) possess the same value in different contexts.

\section{Conclusions}

We have here constructed \textit{conjugate logic}, that includes not only an extension to standard logic introducing an indeterminate ``?'' truth value, but also several conjugate degrees of freedom for each individual system.
Crucially, it also includes transformations between propositions that concern different degrees of freedom, and transformations between single-system propositions and correlation propositions.
The construction is intended to capture not only quantum uncertainty but more properly quantum properties of the described systems, for which the transformations are crucial.

The introduction of a limitation in predictive power, and the implied action of measurements, does make the logic contain quantum-like elements.
But it is really the choice of Hadamard transformation and correlating transformation ($CZ$) that decide if the behavior is quantum or merely quantum-like.
From the point of view of conjugate logic, it is these transformations that are responsible for the emergence of quantum contextuality.

In contrast to quantum logic, the approach presented here does not force contextuality into the logic through postulating Hilbert space structure.
It is instead a generic framework that allows contextuality to occur, or not occur, as desired. 
We aim to use this simple language to describe other quantum and quantum-like phenomena, and hope that the framework will be useful to others in the same line of work.

\section*{Acknowledgements}
The authors would like to thank Andrei Khrennikov for arranging the long-running Växjö conference series on foundations of physics and probability. 
We believe that the friendly atmosphere and the open-minded discussions of the conference series to a large extent is because of Andrei's influence, we have all benefited greatly from participation.

FH acknowledges support by the Government of Spain (FIS2020-TRANQI and Severo Ochoa CEX2019-000910-S), Fundació Mir-Puig, Generalitat de Catalunya (AGAUR SGR 1381 and CERCA), the European Union under Horizon2020 (PROBIST 754510), and the Foundation for Polish Science through TEAM-NET (POIR.04.04.00-00-17C1/18-00).

\printbibliography
\appendix
\section{Truth tables for the logic equivalences and logic implications}
\label{sec:tables}

Some equivalence laws are very simple to check, such as (E1) and (E3)--(E8), and the inverse law (E9) is contained in the main text. What remains are the following. 
\begin{table}[H]
\caption{%Truth table for 
(E2) de Morgan's laws: $\neg(p\wedge q)\Leftrightarrow\neg p\vee\neg q$; $\neg(p\vee q)\Leftrightarrow\neg p\wedge\neg q$}
\label{tab:demorgan}
\centering
\begin{tabular}{@{\hspace{2mm}}c@{\hspace{2mm}}c@{\hspace{5mm}}c@{\hspace{3mm}}c
@{\hspace{5mm}}c@{\hspace{3mm}}c@{\hspace{2mm}}}
\hline
\noalign{\smallskip}
$p$&$q$&$p\wedge q$&$\neg p\vee\neg q$&$p\vee q$&$\neg p\wedge\neg q$\\
\noalign{\smallskip}
\svhline
\noalign{\smallskip}
0&0&0&1&0&1\\
0&?&0&1&?&?\\
0&1&0&1&1&0\\
?&0&?&?&?&?\\
?&?&?&?&?&?\\
?&1&?&?&1&0\\
1&0&0&1&1&0\\
1&?&?&?&1&0\\
1&1&1&0&1&0\\
\noalign{\smallskip}
\hline
\noalign{\smallskip}
\end{tabular}
\end{table}

\begin{table}[H]
\caption{%Truth table for 
(E10) absorption laws: $p\vee(p\wedge q)\Leftrightarrow p$; $p\wedge(p\vee q)\Leftrightarrow p$ }
\label{tab:absorption}
\centering
\begin{tabular}{@{\hspace{2mm}}c@{\hspace{2mm}}c@{\hspace{5mm}}c@{\hspace{3mm}}c
@{\hspace{5mm}}c@{\hspace{3mm}}c@{\hspace{2mm}}}
\hline
\noalign{\smallskip}
$p$&$q$&$p\wedge q$&$p\vee(p\wedge q)$&$p\vee q$&$p\wedge(p\vee q)$\\
\noalign{\smallskip}
\svhline
\noalign{\smallskip}
0&0&0&0&0&0\\
0&?&0&0&?&0\\
0&1&0&0&1&0\\
?&0&0&?&?&?\\
?&?&?&?&?&?\\
?&1&?&?&1&?\\
1&0&0&1&1&1\\
1&?&?&1&1&1\\
1&1&1&1&1&1\\
\noalign{\smallskip}
\hline
\noalign{\smallskip}
\end{tabular}
\end{table}

\begin{table}[H]
\caption{%Truth table for 
(E11) \sout{the implication law}: $(p\rightarrow q)\nLeftrightarrow(\neg p\vee q)$}
\label{tab:implication}
\centering
\begin{tabular}{@{\hspace{2mm}}c@{\hspace{2mm}}c@{\hspace{3mm}}c@{\hspace{3mm}}c
@{\hspace{3mm}}c@{\hspace{3mm}}c@{\hspace{2mm}}}
\hline
\noalign{\smallskip}
$p$&$q$&$p\rightarrow q$&$\neg p\vee q$&$(p\rightarrow q)\rightarrow(\neg p\vee q)$&$(p\rightarrow q)\leftarrow(\neg p\vee q)$\\
\noalign{\smallskip}
\svhline
\noalign{\smallskip}
0&0&1&1&1&1\\
0&?&1&1&1&1\\
0&1&1&1&1&1\\
?&0&0&?&1&?\\
?&?&1&?&?&1\\
?&1&1&1&1&1\\
1&0&0&0&1&1\\
1&?&0&?&1&?\\
1&1&1&1&1&1\\
\noalign{\smallskip}
\hline
\noalign{\smallskip}
\end{tabular}
\end{table}

\begin{table}[H]
\caption{%Truth table for 
(E12) the contrapositive law: $p\rightarrow q\Leftrightarrow\neg q\rightarrow\neg p$}
\label{tab:contrapositive}
\centering
\begin{tabular}{@{\hspace{2mm}}c@{\hspace{2mm}}c@{\hspace{3mm}}c@{\hspace{3mm}}c
@{\hspace{2mm}}c@{\hspace{3mm}}c@{\hspace{2mm}}}
\hline
\noalign{\smallskip}
$p$&$q$&$p\rightarrow q$&$\neg p$&$\neg q$&$\neg q\rightarrow\neg p$\\
\noalign{\smallskip}
\svhline
\noalign{\smallskip}
0&0&1&1&1&1\\
0&?&1&1&?&1\\
0&1&1&1&0&1\\
?&0&0&?&1&0\\
?&?&1&?&?&1\\
?&1&1&?&0&1\\
1&0&0&0&1&0\\
1&?&0&0&?&0\\
1&1&1&0&0&1\\
\noalign{\smallskip}
\hline
\noalign{\smallskip}
\end{tabular}
\end{table}

\begin{table}[H]
\caption{%Truth table for 
(E13) the equivalence law: $p\leftrightarrow q\Leftrightarrow (p\rightarrow q)\wedge(p\leftarrow q)$}
\label{tab:equivalence}
\centering
\begin{tabular}{@{\hspace{2mm}}c@{\hspace{2mm}}c@{\hspace{5mm}}c@{\hspace{5mm}}c
@{\hspace{3mm}}c@{\hspace{2mm}}}
\hline
\noalign{\smallskip}
$p$&$q$&$p\leftrightarrow q$&$p\rightarrow q$&$p\leftarrow q$\\
\noalign{\smallskip}
\svhline
\noalign{\smallskip}
0&0&1&1&1\\
0&?&0&1&0\\
0&1&0&1&0\\
?&0&0&0&1\\
?&?&1&1&1\\
?&1&0&1&0\\
1&0&0&0&1\\
1&?&0&0&1\\
1&1&1&1&1\\
\noalign{\smallskip}
\hline
\noalign{\smallskip}
\end{tabular}
\end{table}

\begin{table}[H]
\caption{%Truth table for 
(I1) Modus ponens: $(p\rightarrow q)\wedge p\Rightarrow q$}
\label{tab:modusponens}
\centering
\begin{tabular}{@{\hspace{2mm}}c@{\hspace{2mm}}c@{\hspace{5mm}}c@{\hspace{3mm}}c
@{\hspace{3mm}}c@{\hspace{2mm}}}
\hline
\noalign{\smallskip}
$p$&$q$&$p\rightarrow q$&$(p\rightarrow q)\wedge p$&$[(p\rightarrow q)\wedge p]\rightarrow q$\\
\noalign{\smallskip}
\svhline
\noalign{\smallskip}
0&0&1&0&1\\
0&?&1&0&1\\
0&1&1&0&1\\
?&0&0&0&1\\
?&?&1&?&1\\
?&1&1&?&1\\
1&0&0&0&1\\
1&?&0&0&1\\
1&1&1&1&1\\
\noalign{\smallskip}
\hline
\noalign{\smallskip}
\end{tabular}
\end{table}

\begin{table}[H]
\caption{%Truth table for 
(I2) law of syllogism: $(p\rightarrow q)\wedge (q\rightarrow r)\Rightarrow p\rightarrow r$}
\label{tab:syllogism}
\centering
\begin{tabular}{@{\hspace{2mm}}c@{\hspace{2mm}}c@{\hspace{2mm}}c@{\hspace{5mm}}c
@{\hspace{3mm}}c@{\hspace{5mm}}c@{\hspace{3mm}}c@{\hspace{2mm}}}
\hline
\noalign{\smallskip}
$p$&$q$&$r$&$p\rightarrow q$&$q\rightarrow r$&$(p\rightarrow q)\wedge(q\rightarrow r)$&$p\rightarrow r$\\
\noalign{\smallskip}
\svhline
\noalign{\smallskip}
0&0&0&1&1&1&1\\
0&0&?&1&1&1&1\\
0&0&1&1&1&1&1\\
0&?&0&1&0&0&1\\
0&?&?&1&1&1&1\\
0&?&1&1&1&1&1\\
0&1&0&1&0&0&1\\
0&1&?&1&0&0&1\\
0&1&1&1&1&1&1\\
?&0&0&0&1&0&0\\
?&0&?&0&1&0&1\\
?&0&1&0&1&0&1\\
?&?&0&1&0&0&0\\
?&?&?&1&1&1&1\\
?&?&1&1&1&1&1\\
?&1&0&1&0&0&0\\
?&1&?&1&0&0&1\\
?&1&1&1&1&1&1\\
1&0&0&0&1&0&0\\
1&0&?&0&1&0&0\\
1&0&1&0&1&0&0\\
1&?&0&0&0&0&0\\
1&?&?&0&1&0&0\\
1&?&1&0&1&0&0\\
1&1&0&1&0&0&1\\
1&1&?&1&0&0&1\\
1&1&1&1&1&1&1\\
\noalign{\smallskip}
\hline
\noalign{\smallskip}
\end{tabular}
\end{table}
\vspace{-1em}

\begin{table}[H]
\caption{%Truth table for 
(I3) Modus tollens: $(p\rightarrow q)\wedge \neg q\Rightarrow \neg p$}
\label{tab:modustollens}
\centering
\begin{tabular}{@{\hspace{2mm}}c@{\hspace{2mm}}c@{\hspace{5mm}}c@{\hspace{3mm}}c
@{\hspace{3mm}}c@{\hspace{2mm}}}
\hline
\noalign{\smallskip}
$p$&$q$&$p\rightarrow q$&$(p\rightarrow q)\wedge\neg q$&$[(p\rightarrow q)\wedge \neg q]\rightarrow \neg p$\\
\noalign{\smallskip}
\svhline
\noalign{\smallskip}
0&0&1&1&1\\
0&?&1&?&1\\
0&1&1&0&1\\
?&0&0&0&1\\
?&?&1&?&1\\
?&1&1&0&1\\
1&0&0&0&1\\
1&?&0&0&1\\
1&1&1&0&1\\
\noalign{\smallskip}
\hline
\noalign{\smallskip}
\end{tabular}
\end{table}
\vspace{-1em}

\begin{table}[H]
\caption{%Truth table for 
(I4) Conjunctive simplification: $p\wedge q\Rightarrow p$; and (I5) Disjunctive strengthening:  $p\Rightarrow p\vee q$ }
\label{tab:conjunctive}
\centering
\begin{tabular}{@{\hspace{2mm}}c@{\hspace{2mm}}c@{\hspace{5mm}}c@{\hspace{3mm}}c
@{\hspace{5mm}}c@{\hspace{3mm}}c@{\hspace{2mm}}}
\hline
\noalign{\smallskip}
$p$&$q$&$p\wedge q$&$(p\wedge q)\rightarrow p$&$p\vee q$&$p\rightarrow(p\vee q)$\\
\noalign{\smallskip}
\svhline
\noalign{\smallskip}
0&0&0&1&0&1\\
0&?&0&1&?&1\\
0&1&0&1&1&1\\
?&0&0&1&?&1\\
?&?&?&1&?&1\\
?&1&?&1&1&1\\
1&0&0&1&1&1\\
1&?&?&1&1&1\\
1&1&1&1&1&1\\
\noalign{\smallskip}
\hline
\noalign{\smallskip}
\end{tabular}
\end{table}

\begin{table}[H]
\caption{%Truth table for 
(I6) \sout{Disjunctive syllogism}: $(p\vee q)\wedge\neg q\nRightarrow p$}
\label{tab:disjunctive}
\centering
\begin{tabular}{@{\hspace{2mm}}c@{\hspace{2mm}}c@{\hspace{5mm}}c@{\hspace{2mm}}c
@{\hspace{5mm}}c@{\hspace{2mm}}c@{\hspace{2mm}}}
\hline
\noalign{\smallskip}
$p$&$q$&$p\vee q$&$(p\vee q)\wedge\neg q$&$[(p\vee q)\wedge\neg q]\rightarrow p$\\
\noalign{\smallskip}
\svhline
\noalign{\smallskip}
0&0&0&0&1\\
0&?&?&?&?\\
0&1&1&0&1\\
?&0&?&?&1\\
?&?&?&?&1\\
?&1&1&0&1\\
1&0&1&1&1\\
1&?&1&?&1\\
1&1&1&0&1\\
\noalign{\smallskip}
\hline
\noalign{\smallskip}
\end{tabular}
\end{table}

\begin{table}[H]
\caption{%Truth table for 
(I7) proof by contradiction: $(\neg p\rightarrow \gen{\neg I})\Rightarrow p$}
\label{tab:contradiction}
\centering
\begin{tabular}{@{\hspace{2mm}}c@{\hspace{3mm}}c@{\hspace{3mm}}c@{\hspace{2mm}}}
\hline
\noalign{\smallskip}
$p$&$\neg p\rightarrow \gen{\neg I}$&$(\neg p\rightarrow \gen{\neg I})\rightarrow p$\\
\noalign{\smallskip}
\svhline
\noalign{\smallskip}
0&0&1\\
?&0&1\\
1&1&1\\
\noalign{\smallskip}
\hline
\noalign{\smallskip}
\end{tabular}
\end{table}

%(I8)&\textbf{Proof by cases}&
\begin{table}[H]
\caption{%Truth table for 
(I8) proof by cases: $(p\rightarrow r)\wedge (q\rightarrow r)\Rightarrow (p\vee q)\rightarrow r$}
\label{tab:cases}
\centering
\begin{tabular}{@{\hspace{2mm}}c@{\hspace{2mm}}c@{\hspace{2mm}}c@{\hspace{5mm}}c
@{\hspace{3mm}}c@{\hspace{3mm}}c@{\hspace{3mm}}c@{\hspace{2mm}}}
\hline
\noalign{\smallskip}
$p$&$q$&$r$&$p\rightarrow r$&$q\rightarrow r$&$(p\rightarrow r)\wedge (q\rightarrow r)$&$(p\vee q)\rightarrow r$\\
\noalign{\smallskip}
\svhline
\noalign{\smallskip}
0&0&0&1&1&1&1\\
0&0&?&1&1&1&1\\
0&0&1&1&1&1&1\\
0&?&0&1&0&0&0\\
0&?&?&1&1&1&1\\
0&?&1&1&1&1&1\\
0&1&0&1&0&0&0\\
0&1&?&1&0&0&0\\
0&1&1&1&1&1&1\\
?&0&0&0&1&0&0\\
?&0&?&1&1&1&1\\
?&0&1&1&1&1&1\\
?&?&0&0&0&0&0\\
?&?&?&1&1&1&1\\
?&?&1&1&1&1&1\\
?&1&0&0&0&0&0\\
?&1&?&1&0&0&0\\
?&1&1&1&1&1&1\\
1&0&0&0&1&0&0\\
1&0&?&0&1&0&0\\
1&0&1&1&1&1&1\\
1&?&0&0&0&0&0\\
1&?&?&0&1&0&0\\
1&?&1&1&1&1&1\\
1&1&0&0&0&0&0\\
1&1&?&0&0&0&0\\
1&1&1&1&1&1&1\\
\noalign{\smallskip}
\hline
\noalign{\smallskip}
\end{tabular}
\end{table}

\section{Choice of \textit{CZ} transformation}
\label{sec:inconsistency}

This appendix contains a derivation of an inconsistency that would arise if the $CZ$ transformation is chosen so that
\begin{equation}
\widetilde\varphi_{CZ} \gen{XX} \Leftrightarrow \gen{\neg YY},
\end{equation}
where the tilde is used to distinguish the transformation used in the main text from the one considered here. 
The inconsistency arises when making predictions from the conjunction $\gen{\neg YYI,\neg IYY}$.
To generate predictions we first use the procedure of Section~\ref{sec:predictions}.
This starts with the simultaneous Clifford reduction
\begin{equation}
\begin{split}
\gen{\neg YYI,\neg IYY}
&\Leftrightarrow
\varphi_{SSS}\gen{\neg XXI,\neg IXX}
\\&\Leftrightarrow
\varphi_{SSS}\varphi_{IHH}\gen{\neg XZI,\neg IZZ}
\\&\Leftrightarrow
\varphi_{SSS}\varphi_{IHH}\widetilde\varphi_{CZ_{12}}\gen{\neg XII,\neg IZZ}
\\&\Leftrightarrow
\varphi_{SSS}\varphi_{IHH}\widetilde\varphi_{CZ_{12}}\varphi_{IHI}\gen{\neg XII,\neg IXZ}
\\&\Leftrightarrow
\varphi_{SSS}\varphi_{IHH}\widetilde\varphi_{CZ_{12}}\varphi_{IHI}\widetilde\varphi_{CZ_{23}}\gen{\neg XII,\neg IXI}.
\label{eq:cliffordreduction}
\end{split}
\end{equation}
We have the simple prediction
\begin{equation}
\gen{\neg XII,\neg IXI}\Rightarrow\gen{XXI}.
\end{equation}
Clifford expansion now gives
\begin{equation}
\begin{split}
\varphi_{SSS}^{-1}&\varphi_{IHH}\widetilde\varphi_{CZ_{12}}\varphi_{IHI}\widetilde\varphi_{CZ_{23}}\gen{XXI}\\
&\Leftrightarrow
\varphi_{SSS}^{-1}\varphi_{IHH}\widetilde\varphi_{CZ_{12}}\varphi_{IHI}\gen{XXZ}
\\&\Leftrightarrow
\varphi_{SSS}^{-1}\varphi_{IHH}\widetilde\varphi_{CZ_{12}}\gen{XZZ}
\\&\Leftrightarrow
\varphi_{SSS}^{-1}\varphi_{IHH}\gen{XIZ}
\\&\Leftrightarrow
\varphi_{SSS}^{-1}\gen{XIX}
\\&\Leftrightarrow
\gen{YIY},
\end{split}
\end{equation}
which gives the prediction
\begin{equation}
\gen{\neg YYI,\neg IYY}\Rightarrow\gen{YIY}.
\label{eq:YIYpredictiona}
\end{equation}
A second prediction from $\gen{\neg YYI,\neg IYY}$ starts from
\begin{equation}
\gen{XXI,IXX}\Rightarrow\gen{XIX}, 
\end{equation}
that can also be obtained as above.
We now apply three $\widetilde\varphi_{CZ}$ transformations to reach
\begin{equation}
\begin{split}
\gen{\neg YYI,\neg IYY}
&\Leftrightarrow
\widetilde\varphi_{CZ_{12}}\widetilde\varphi_{CZ_{23}}\widetilde\varphi_{CZ_{13}}\gen{XXI,IXX}\\
&\Rightarrow
\widetilde\varphi_{CZ_{12}}\widetilde\varphi_{CZ_{23}}\widetilde\varphi_{CZ_{13}}\gen{XIX}
\Leftrightarrow
\gen{\neg YIY}.
\end{split}
\label{eq:YIYpredictionb}
\end{equation}
Using $\widetilde\varphi_{CZ}$ results in two predictions \eqref{eq:YIYpredictiona} and \eqref{eq:YIYpredictionb}, so that $\gen{\neg YYI,\neg IYY}\Rightarrow\gen{YIY, \neg YIY}$.
While $\gen{YIY,\neg YIY}$ is never true, the Clifford reduction \eqref{eq:cliffordreduction} tells us that the left-hand side $\gen{\neg YYI,\neg IYY}$ can be reduced to compatible propositions for two different systems, and therefore can be simultaneously true.
This shows that $\widetilde\varphi_{CZ}$ gives an inconsistency.

Using $\varphi_{CZ}$ instead of $\widetilde\varphi_{CZ}$ does not change the prediction in Eqn.~\eqref{eq:YIYpredictiona}, but gives
\begin{equation}
\begin{split}
\gen{\neg YYI,\neg IYY}
&\Leftrightarrow
\varphi_{CZ_{12}}\varphi_{CZ_{23}}\varphi_{CZ_{13}}\gen{\neg XXI,\neg IXX}\\
&\Rightarrow
\varphi_{CZ_{12}}\varphi_{CZ_{23}}\varphi_{CZ_{13}}\gen{XIX}
\Leftrightarrow
\gen{YIY}.
\end{split}
\end{equation}
For this choice there is no inconsistency.
\end{document}